\documentclass[aps,prb,twocolumn,groupedaddress,showpacs,floatfix]{revtex4}
\usepackage{graphicx}

\begin{document}

\title{Study of charge dynamics in transparent single-walled carbon nanotube films}


\author{F. Borondics}
%\email[]{bf@szfki.hu}
%\thanks{}
%\affiliation{}
\author{K. Kamar\'as}
\email[]{kamaras@szfki.hu}
%\thanks{}
\affiliation{Research Institute for Solid State Physics and
Optics, Hungarian Academy of Sciences, P.O. Box 49, Budapest,
Hungary H 1525}
\author{M. Nikolou}
\thanks{Present address: Cornell Center for Materials Research,
Cornell University, Ithaca, NY 14853, U.S.A.}
%\affiliation{}
\author{D.B. Tanner}
%\thanks{}
%\affiliation{}
\author{Z.H. Chen}
\thanks{Present address: IBM T.J. Watson Center, P.O. Box 218, Yorktown Heights, NY 10598, U.S.A.}
%\affiliation{}
\author{A.G. Rinzler}
%\thanks{}
\affiliation{Department of Physics, University of Florida,
Gainesville, FL 32611, U.S.A.}

\date{\today}

\begin{abstract}
We report the transmission over a wide frequency range (far
infrared--visible) of pristine and hole-doped, free-standing carbon
nanotube films at
temperatures between 50 K and 300 K. Optical constants are estimated
by Kramers-Kronig analysis of transmittance.  We see evidence in the far
infrared for a gap below 10 meV.  Hole doping causes a shift of
spectral weight from the first interband transition into the far
infrared. Temperature dependence in both the doped and
undoped samples is restricted to the far-infrared region.
\end{abstract}


\maketitle


\section{Introduction}

As carbon nanotubes attract more and more attention for use as electronic materials, so grows the need for the
accurate determination of their fundamental electrical and optical properties. However, for practical reasons,
these properties are not easy to determine. The optical response of a carbon nanotube depends on the tube
diameter, chirality, and orientation. Bulk samples, as well as thin films, are made up of many tube types which
differ in the values of these parameters. In addition the samples have a rough surface, contain void space and
have variability regarding the interaction between tubes, \emph{e.g.} the strength of intertube hopping. Despite
these less than ideal properties, applications exist where single-walled carbon nanotube (SWNT) thin films offer
superior performance to other materials.\cite{wu04} In this paper we present the frequency-dependent optical
functions determined from transmission through high-quality free-standing nanotube films in a wide frequency and
temperature range, followed by Kramers-Kronig analysis of transmittance.

Nanotube optical properties have been studied by a number of researchers. Some studies
\cite{kataura99,oconnell02} have been restricted to photon energies between 0.5--3 eV. Other
measurements\cite{ugawa99,ugawa01,itkis02} have included lower frequencies, where  metallic tubes exhibit
free-carrier absorption. Studies in the low frequency range also allow one to test the prediction that certain
tubes would have very small gaps.\cite{kane97,delaney98,maarouf00} These gaps have been reported near 10 meV,
with substantial variation from sample to sample.\cite{ugawa99,ugawa01,itkis02}

Our measurements improve on earlier reports\cite{ugawa99,hwang00,ugawa01,itkis02,hennrich02,hennrich03} in
several ways. Thick films used for reflectance measurements\cite{ugawa99,hwang00,ugawa01} had poor surface
quality. Transmission of films deposited on substrates\cite{itkis02} required the use of several different
substrate materials, making normalization difficult; moreover, temperature studies in this configuration are
hindered by the thermal properties of the substrates. Transmission studies\cite{hennrich02} of free-standing
nanotube films found evidence for free-carrier response down to 50 meV but were not analyzed for optical
constants. We present here measurements over a wide frequency range on free-standing samples, studying both the
temperature dependence and the redistribution of carriers upon doping.

\section{Experimental procedures}

\begin{table}%[H] add [H] placement to break table across pages
\caption {Processing details of the samples studied.
\label{table:proc}}
\begin{ruledtabular}
\begin{tabular}{lcccc}
ID & Thickness(nm)  & \multicolumn{2}{c}{Processing} &  \\
   &                & temperature ($^{\circ}$C) & time (h) & \\
\hline
A &  250 & as-prepared & & \\
B &  250 & 1000  & 12  & \\
C &  150 & as-prepared & & \\
C{$^\prime$} & 150 & 1000  & 12  & \\
\end{tabular}
\end{ruledtabular}
\end{table}

Film preparation, starting from laser-ablated SWNT's of 1.37 nm mean diameter,\cite{rinzler98} was described in
Ref.~\onlinecite{wu04}. Details of processing for the samples are given in Table \ref{table:proc}. The sample
thicknesses were determined by AFM measurements. Samples A and C were measured as prepared (``unbaked''). Thermal
processing (``baking'') consisted of treating the samples in flowing pure argon at 1000$^\circ$~C for 12 hours,
to remove from the sample the nitric acid, used during purification.\cite{itkis02,hennrich03} Sample B was a
second piece of the same film as Sample A; samples C and C{$^\prime$} were the same piece of film measured both
before and after baking. Spectra were taken on three different spectrometers with sufficient spectral overlap to
allow unambiguous merging of transmittance curves: a Bruker 113v Fourier-transform interferometer (20--3000
cm$^{-1}$) with a Si bolometer (FIR) and a DTGS detector (MIR), a home-made near infrared/visible/UV spectrometer
based on a Perkin-Elmer monochromator (2000--40000 cm$^{-1}$), and a Perkin-Elmer Lambda 900 VIS/UV spectrometer.
Temperature-dependent measurements were conducted with the first two instruments in a flowthrough He cryostat
with polyethylene and KBr windows, respectively. The polyethylene windows caused interference fringes in the
transmission spectra which, however, do not change the level of transmission. We used a transparent sample of
5~mm diameter, producing enough signal even at the lowest frequencies where transmission was low. The spectra
were reproducible during several cooling and heating cycles.

In the usual analysis of visible/near-IR spectra, one calculates the extinction (\emph{-ln~T} or \emph{-log~T})
from the transmission and calls this quantity the absorbance, or the product of absorption coefficient times
thickness, without making further corrections. For many materials this method is perfectly adequate, as the
amount of light reflected by the sample is small and only weakly dependent on wavelength. However, this procedure
is not justified for strongly absorbing materials; in these the reflection (and its wavelength dependence) is a
major factor, indeed often the dominant one,  in determining the light transmission. This situation is the case
in nanotube samples at low frequencies. Therefore, we determined the optical functions by Kramers-Kronig analysis
of the transmittance.

Kramers-Kronig analysis is not as commonly applied to transmittance
as it is to reflectance. Nevertheless, the transmittance of a film
is subject to the same causality restrictions as the reflectance;
consequently, one may estimate the phase shift on transmittance from
a Kramers-Kronig integral, much as one does for reflectance. The
procedure requires free-standing, uniform-thickness films,  wide
spectral coverage, and reasonable photometric accuracy,
${\cal O}(1\%)$. The relationship between the phase shift
$\Theta(\omega)$ and the transmittance $T(\omega)$
may be written in the following illuminating form:
\begin{equation}\label{KK1}
\Theta(\omega) =
-\frac{1}{2\pi}\int_0^{\infty}\ln {\vert
{s+\omega}\vert\over\vert{s-\omega}\vert} {d\ln T(s)\over ds}ds.
\end{equation}
According to Equation~\ref{KK1}, spectral regions in which the transmittance
is constant do not contribute to the integral. Further, the spectral
regions $s\gg \omega$ and $s\ll \omega$ do not contribute much because the
function $\ln \vert (s+\omega)/(s-\omega)\vert$ is small in these regions.

Formally, the phase-shift integral requires knowledge of the
transmittance at all frequencies. In practice, one obtains the
transmittance over as wide a frequency range as possible and then
completes the transform by extrapolating the transmittance to
frequencies above and below the range of the available measurements.
The conventional low-frequency extrapolation for metals is
 $T(\omega) = T(0) + A\omega^2$, where $A$ is
a constant determined by the transmittance at the lowest frequency
measured in the experiment and $T(0)$ is the extrapolated behavior to
zero frequency, determined  by the dc conductivity. The high-frequency
extrapolation uses $T = 1 - C\omega^{-n}$ with $n\approx 1$ and $C$ chosen to give a smooth
connection to the high-frequency transmittance curve.

After computing the phase, one may extract the complex refractive
index $N$ (and all other optical constants) by numerical solution of\cite{dresselgruner02}
\begin{equation}\label{KK2}
    \sqrt{T}e^{i\Theta}=\frac{4N}{(N+1)^2e^{-i\delta}-(N-1)^2e^{i\delta}},
\end{equation}
where $\delta={\omega N d}/{c}$ with $d$ the film thickness. An important detail is
that the phase gained by the radiation in passing through a
thickness $d$ of vacuum must be added to $\delta$  before calculating
$N$.
Equation \ref{KK2} includes of course interfacial
reflections, including multiple reflections within the sample.

\section{Results}

\begin{figure}%[htbp]
\includegraphics[]{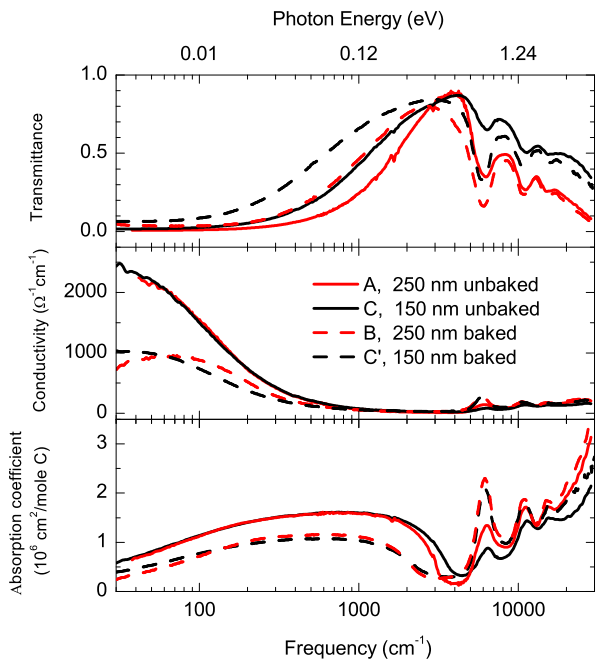}
\caption{(color online) Transmittance, optical conductivity and absorption coefficient of two SWNT
transparent films of different thickness.
\label{fig:sampdep}}
\end{figure}

Figure~\ref{fig:sampdep} shows the room-temperature transmission, optical conductivity and calculated absorbance
spectra before and after baking. Above 4000 cm$^{-1}$, the absorption coefficients calculated from the
Kramers-Kronig transformation agree with (\emph{-ln T}).

The absorption coefficient
of the material is the absorbance per unit
thickness; to obtain absorption coefficients in units of cm$^2/$(mole
C), we have to divide by the density. We determined the density of
our materials directly by weighing a piece whose thickness was
measured by atomic force microscopy, and obtained 0.4~g/cm$^3$.
 Our absorption values are higher than those found
by Zhao {\it et al.\/}\cite{zhao04} in a comprehensive study of
various nanotubes. They found $\alpha(S_{22}$) of $0.6$--$1.1\cdot
10^6$ (in our units), compared with our value (in the baked
samples) of $1.8$--$1.9\cdot 10^6$. One reason for the difference
may be that in our films the long axes of the tubes are largely
aligned in the plane of the film, whereas in solution spectroscopy
they are disordered in three dimensions.

\begin{figure*}%[htbp]
\includegraphics[]{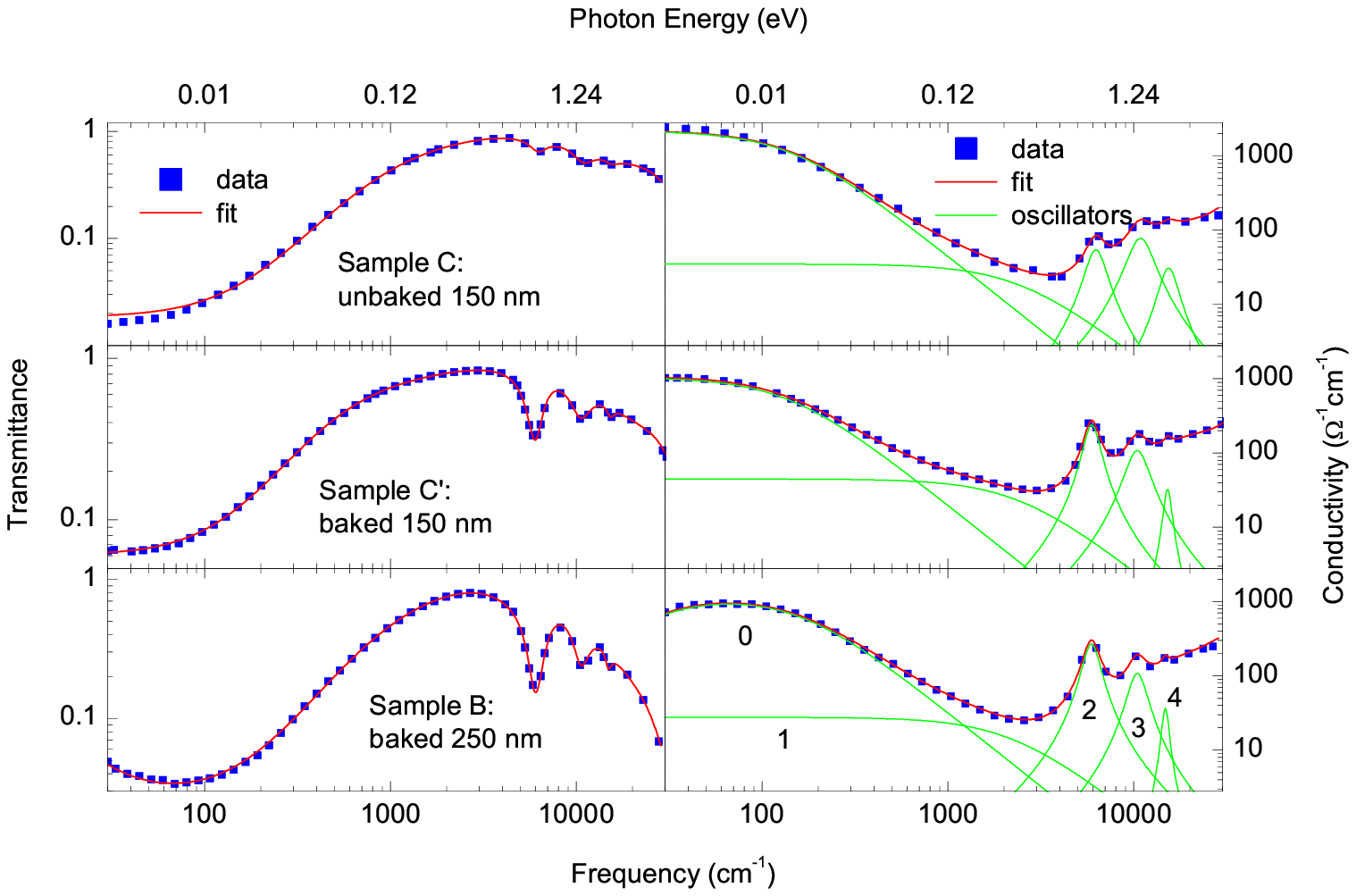}%
\caption{(color online) The results of Drude-Lorentz fits (least-squares fits to transmission curves) compared to
measured room-temperature transmission data and to conductivity obtained by Kramers-Kronig transformation.
Numbers refer to oscillator labels in Table~\ref{table:dlparm}. \label{fig:fits}}
\end{figure*}

We also performed a Drude-Lorentz fit to the transmission curves, using the Airy formula\cite{dresselgruner02}
for the transmittance of a thin, free-standing film. Table \ref{table:dlparm} lists the fitting parameters
obtained. The fits and the contributions of individual oscillators to the optical conductivity at 300 K are
illustrated in Fig. \ref{fig:fits}. We used only electronic oscillators to model the spectra; no vibrational
transitions are discernible in the baked samples, even after the baseline correction described recently by Kim
{\it et al.\/}\cite{Kim05} In the doped materials, we observed the weak vibrational peaks seen by Hennrich {\it
et al.\/};\cite{hennrich03} these disappear on baking.

\begin{table*}%
\caption {Drude-Lorentz fit parameters (plasma frequency ($\omega_{pi}$), center frequency ($\omega_i$)  and
width ($\gamma_i$) in cm$^{-1}$) for each sample studied at several temperatures. For peak 0, the free-carrier
(Drude) absorption, the center frequency is 0. The starting parameters for high-frequency oscillators were taken
from Ref. \onlinecite{murakami05prl}. The high-frequency dielectric constant is 1.04.\label{table:dlparm}}
\begin{ruledtabular}
\begin{tabular}{lrrrrrrrrrrrrrrr}
%\hline
 & & $\omega_{p0}$ & $\gamma_0$ & $\omega_{p1}$ & $\omega_1$ & $\gamma_1$ & $\omega_{p2}$ & $\omega_2$ & $\gamma_2$ & $\omega_{p3}$ & $\omega_3$ & $\gamma_3$ & $\omega_{p4}$ & $\omega_4$ & $\gamma_4$  \\
\hline\hline
% & & & & & & & & & & & & \\
Sample A&300 K & 2057 & 762 & 4055 & 35 & 128 & 3419 & 6189 & 1435 & 5268 & 10616 & 3888 & 2505 & 14884 & 2199 \\
doped  & & & & & & & & & & & & & & \\
\hline
%& & & & & & & & & & & & & & \\
Sample B & 50 K & 1817 & 2555 & 3257 & 58 & 154 & 4613 & 5913 & 1298 & 4105 & 10429 & 2587 & 1689 & 14763 & 1310 \\
de-doped     & 100 K & 1927 & 2661 & 3258 & 56 & 158 & 4615 & 5911 & 1308 & 4079 & 10436 & 2552 & 1721  & 14771 & 1342 \\
          & 200 K & 1863 & 2250 & 3242 & 65 & 166 & 4573 & 5913 & 1277 & 4085 & 10424 & 2597 & 1676 & 14764 & 1306 \\
          & 300 K & 1926 & 2232 & 3226 & 66 & 187 & 4600 & 5910 & 1310 & 4076 & 10435 & 2571 & 1725 & 14772 & 1355 \\
\hline
% & & & & & & & & & & & & & & \\
 Sample C & 100 K & 2314 & 3393 & 4408 & 0 & 134 & 2296 & 6245 & 1648 & 4074 & 10859 & 3591 & 2674 & 15400 & 3880 \\
 most doped &  200 K & 2318 & 2944 & 4381 & 0 & 139 & 2318 & 6241 & 1679 & 4083 & 10868 & 3603 & 2646 & 15402 & 3802 \\
         &  300 K & 2302 & 2509 & 4337 & 0 & 144 & 2328 & 6244 & 1655 & 4114 & 10858 & 3624 & 2646 & 15380 & 3793 \\
\hline
%& & & & & & & & & & & & & & \\
 Sample C{$^\prime$} & 100 K & 1984 & 2136 & 2904 & 20 & 122 & 4044 & 5937 & 1176 & 4459 & 10442 & 3122 & 1937 & 15136 & 1977 \\
 de-doped  & 200 K & 2220 & 2307 & 2917 & 20 & 131 & 4037 & 5919 & 1162 & 4502 & 10410 & 3187 & 1975 & 15194 & 2014 \\
          & 300 K & 2561 & 2442 & 2917 & 20 & 140 & 4206 & 5935 & 1230 & 4520 & 10436 & 3160 & 1756 & 15194 & 1585 \\

\end{tabular}
\end{ruledtabular}
\end{table*}

The conductivity and the absorbance are independent of thickness except at very low frequencies; even there, the
values are comparable.

\section{Discussion}
\subsection{Baked (undoped) samples}

We discuss first the room-temperature results on the baked
samples, because we regard these as closest to pristine carbon
nanotubes. Optical spectroscopy is routinely used for purity
evaluation of nanotube samples\cite{itkis03,zhao04} where the
amount of amorphous carbon is deduced from the background
absorption in the near infrared. Comparing our spectra with those
by Zhao et al.\cite{zhao04} we estimate the purity to equal those
of their best samples. We assume the films dense and isotropic
enough to apply a model of a continous medium. (Effective-medium
models have been employed for polarized measurements where the
shape of the tubes is crucial.\cite{jeon04})

The low-frequency conductivity can be best described by a two-component model. In Sample B, the two components
consist of a Drude free-electron contribution and a Lorentzian peak centered near 0.01 eV. In Sample
C{$^\prime$}, we found that a fit with two Drude contributions, with distinct damping constants, also described
the data. We note, however, that if the center frequency of the Lorentzian shifts below the measured range ($\sim
30$~cm$^{-1}$), then it is essentially indistinguishable from a Drude curve. Indeed, an equally good fit is
obtained if we fix the frequency of this peak at 20 cm$^{-1}$ (close to that observed by Terahertz
spectroscopy\cite{jeon04}). We can also compare our data to those of Hilt \emph{et al.}\cite{hilt00}  measured at
Terahertz frequencies in similar laser-ablated nanotubes as used in the present study. Although the exact doping
level of the tubes studied there is not known, they also find a narrow Drude component and a broad absorption,
modeled as a constant background conductivity in their restricted frequency range.

Thus we assign the Drude part to truly metallic tubes and the
low-frequency Lorentzian to the curvature-induced gap suggested by
Kane and Mele\cite{kane97} and observed in individual nanotubes by
scanning tunneling spectroscopy.\cite{ouyang01} In our samples,
this gap actually represents an average of gaps for
``semimetallic'' tubes of different diameter and
chirality.\cite{ugawafer01} Note that the width ($\gamma$) of the
Drude contribution is typically more than ten times the width of
the low-energy Lorentzian, suggesting that the mobility of
carriers in semiconducting tubes is much higher than in metallic
tubes.

The origin of the variation among samples can be attributed to unintentional doping by atmospheric
oxygen.\cite{sumana00,collins00} A modest number of extra holes is enough to smear out the gaps of some of the
nanotubes and change their distribution, shifting the median to lower energies. The fact that the thinner film
(Sample C{$^\prime$}) appears more doped is in accordance with observations by Collins {\it et
al.\/}\cite{collins00} We see no indication of the bundle-induced pseudogap at 0.1 eV seen in Ref.
\onlinecite{ouyang01}. The disappearing of the pseudogap was predicted by Maarouf {\it et al.\/}\cite{maarouf00}
for macroscopic samples with compositional disorder.

\begin{figure}%[htbp]
\includegraphics[]{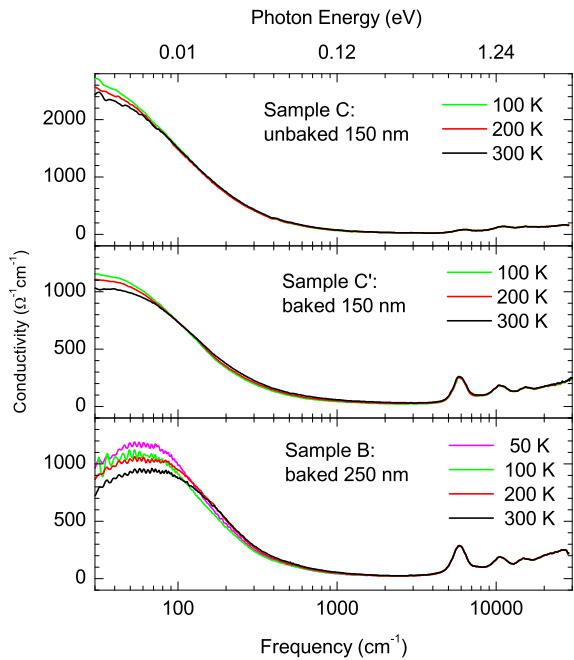}
\caption{(color online) Temperature dependence of the optical conductivity of the two samples. \label{fig:Tdep}}
\end{figure}

The temperature dependence of the conductivity is much weaker than
that of a simple metal (see Fig.~\ref{fig:Tdep}). It is confined
largely to the far infrared. The low-frequency conductivity
decreases with increasing temperature; a crossover at 120
cm$^{-1}$ reverses this dependence up to about 2000 cm$^{-1}$,
where it becomes too weak to be observed. The fits to a dielectric
function model (Table \ref{table:dlparm}) show a slight broadening
of peak 1, along with a blue shift of the low-frequency peak in
Sample B. We saw no indication down to 50 K for a sudden
conductivity drop, which is expected if a Luttinger liquid state
develops.\cite{ishii03}

The low-frequency optical conductivity of the baked samples, 800 and 1000
$\Omega^{-1}$cm$^{-1}$ for Samples B and C$^\prime$, respectively, agree well with
the 700 $\Omega^{-1}$cm$^{-1}$ value determined from sheet
resistance.\cite{wu04} The 2600 $\Omega^{-1}$cm$^{-1}$ dc conductivity of the
unbaked sample is somewhat lower than that determined by direct
sheet resistance measurements (6700 $\Omega^{-1}$cm$^{-1}$)
(and closer to the value 2900 $\Omega^{-1}$cm$^{-1}$ obtained by Liu
{\it et al.\/}\cite{liu04} for FeCl$_3$ doped nanotubes from
EELS results) but the trend of higher
conductivity in the unbaked samples is clear.

\subsection{Unbaked (doped) samples}

We now turn to the effect of doping on the optical properties.
Figure \ref{fig:sampdep} demonstrates that nitric acid doping increases
the conductivity in the far infrared and
decreases it around the first Van Hove transition.
At low frequencies, the narrower band (the one at finite frequency
in Sample B) is much more influenced than the broad Drude part. We
can directly compare the parameters of Sample C (doped) and
C{$^\prime$} (undoped): as a general trend, peak 0 and peak 1 both
increase in intensity on doping, peak 1 more so than peak 0, while
peak 2 and peak 3 (the semiconductor tubes' S$_{11}$ and S$_{22}$
transitions, respectively) decrease. Peak 2 and peak 3 also
shift to higher energies, as already reported by Itkis {\it et
al.\/}\cite{itkis02} The high-frequency changes upon doping have
been reported and explained extensively.\cite{kaza99,petit99} On
hole doping, metallic tubes get depleted of charge and therefore a
slight decrease in their contribution to the low frequency
conductivity is expected.
In semiconductors a redistribution of oscillator
strength occurs  as the highest-lying valence band loses electrons.
This change in the occupation of states leads to a decrease of the
S$_{11}$ interband transition (peak 2) and at the same time, an
increase in the low-frequency conductivity. The fact that this distribution strengthens peak 1
indicates that the doped semiconducting tubes also can be regarded
as semimetals. At higher doping level, the same scenario
occurs at the second interband transition.

\begin{figure}%[htbp]
\includegraphics[]{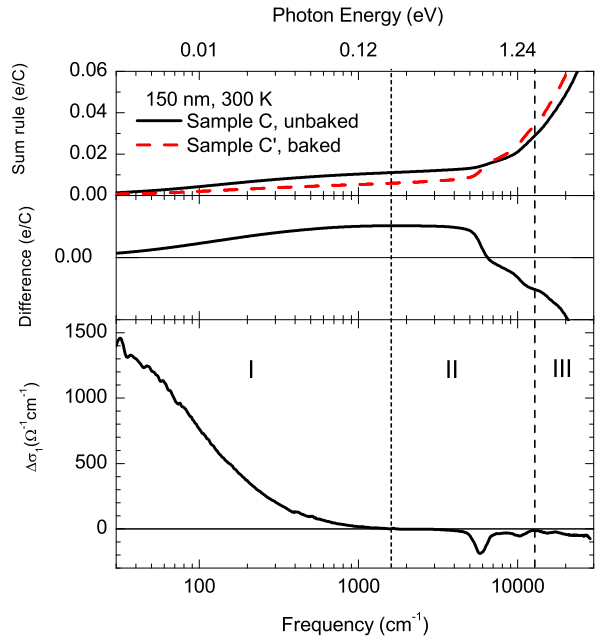}%
\caption{(color online) Top panel: sum rule for baked and unbaked sample at 300 K. Middle panel: added carriers
per carbon as function of frequency at selected temperatures. Bottom panel: difference in conductivity upon
doping. For the definition of regions I, II, and III, see text.
\label{fig:Diffs}}
\end{figure}

The top panel of Fig.~\ref{fig:Diffs} shows the partial sum rule
for Samples C and C{$^\prime$}. This quantity is obtained by
integrating $\sigma_1(\omega)$ from $\omega = 0$ to an upper
frequency, and plotting the integrated weight versus the upper
limit, representing the effective number of carriers per C atom
taking part in optical transitions below the upper-limit
frequency.  Up to the near infrared, the hole-doped material shows
a much higher weight than the de-doped one; at higher frequencies
the difference changes sign. The middle and bottom panels of
Figure \ref{fig:Diffs} show respectively the difference in
spectral weight and the difference in conductivity
$\Delta\sigma_1$ between the doped and undoped samples. The
behavior may be divided into three regions: the free-carrier part
(region I), the interband transition part (region II) and the
$\pi$-electron excitation part (region III). The boundary between
regions I and II is at 1765 cm$^{-1}$, where the conductivity
difference changes sign (bottom panel): we assume that if we had a
series of concentrations, this point would be isosbestic.  The sum
rule difference in this part of the spectrum (middle panel) levels
off at the value of 0.005 e/C, before it starts to decrease at the
S$_{11}$ transition. A second, smaller, negative peak in the
conductivity can be seen at the S$_{22}$ frequency. We put the
boundary between regions II and III at 12,600 cm$^{-1}$, where
$\Delta\sigma_1$ approaches zero again, and the value of the sum
rule reaches -0.005 e/C. We infer from this that the principal
effect of hole doping is a redistribution of charge within the
semiconducting nanotubes, and the doping of the metallic tubes is
negligible in comparison. In region III the collective excitations
of $\pi$-electrons can be found.\cite{pichler99}

\section{Summary}

In summary, we have presented optical transmission in a wide frequency
and temperature range of high-quality transparent single-walled carbon
nanotube films. We found a gap feature below 10 meV and associate it
with the curvature-induced gap of semimetallic tubes. We estimate
the hole doping of our samples by nitric acid to cause a 0.005 e/C
redistribution of charge between the free-carrier absorption and the
interband transitions in semiconducting tubes and suggest that the
free carriers in the doped semiconductors behave similarly to those
in semimetallic tubes.

\begin{acknowledgments}
We thank T. Pichler for enlightening discussions. Work supported by OTKA
T049338 and jointly by NSF-DMR-0305043 and DOE
DE-AI02-03ER46070.
\end{acknowledgments}

\bibliography{nano}

\begin{thebibliography}{29}
\expandafter\ifx\csname natexlab\endcsname\relax\def\natexlab#1{#1}\fi
\expandafter\ifx\csname bibnamefont\endcsname\relax
  \def\bibnamefont#1{#1}\fi
\expandafter\ifx\csname bibfnamefont\endcsname\relax
  \def\bibfnamefont#1{#1}\fi
\expandafter\ifx\csname citenamefont\endcsname\relax
  \def\citenamefont#1{#1}\fi
\expandafter\ifx\csname url\endcsname\relax
  \def\url#1{\texttt{#1}}\fi
\expandafter\ifx\csname urlprefix\endcsname\relax\def\urlprefix{URL }\fi
\providecommand{\bibinfo}[2]{#2}
\providecommand{\eprint}[2][]{\url{#2}}

\bibitem[{\citenamefont{Wu et~al.}(2004)\citenamefont{Wu, Chen, Du, Logan,
  Sippel, Nikolou, Kamar\'as, Reynolds, Tanner, Hebard, and Rinzler}}]{wu04}
\bibinfo{author}{\bibfnamefont{Z.}~\bibnamefont{Wu}},
  \bibinfo{author}{\bibfnamefont{Z.}~\bibnamefont{Chen}},
  \bibinfo{author}{\bibfnamefont{X.}~\bibnamefont{Du}},
  \bibinfo{author}{\bibfnamefont{J.}~\bibnamefont{Logan}},
  \bibinfo{author}{\bibfnamefont{J.}~\bibnamefont{Sippel}},
  \bibinfo{author}{\bibfnamefont{M.}~\bibnamefont{Nikolou}},
  \bibinfo{author}{\bibfnamefont{K.}~\bibnamefont{Kamar\'as}},
  \bibinfo{author}{\bibfnamefont{J.~R.} \bibnamefont{Reynolds}},
  \bibinfo{author}{\bibfnamefont{D.~B.} \bibnamefont{Tanner}},
  \bibinfo{author}{\bibfnamefont{A.~F.} \bibnamefont{Hebard}}, \bibnamefont{and}
  \bibinfo{author}{\bibfnamefont{A.~G.} \bibnamefont{Rinzler}},
  \bibinfo{journal}{Science}
  \textbf{\bibinfo{volume}{305}}, \bibinfo{pages}{1273} (\bibinfo{year}{2004}).

\bibitem[{\citenamefont{Kataura et~al.}(1999)\citenamefont{Kataura, Kumazawa,
  Maniwa, Umezu, Suzuki, Ohtsuka, and Achiba}}]{kataura99}
\bibinfo{author}{\bibfnamefont{H.}~\bibnamefont{Kataura}},
  \bibinfo{author}{\bibfnamefont{J.}~\bibnamefont{Kumazawa}},
  \bibinfo{author}{\bibfnamefont{Y.}~\bibnamefont{Maniwa}},
  \bibinfo{author}{\bibfnamefont{I.}~\bibnamefont{Umezu}},
  \bibinfo{author}{\bibfnamefont{S.}~\bibnamefont{Suzuki}},
  \bibinfo{author}{\bibfnamefont{Y.}~\bibnamefont{Ohtsuka}}, \bibnamefont{and}
  \bibinfo{author}{\bibfnamefont{Y.}~\bibnamefont{Achiba}},
  \bibinfo{journal}{Synth.\ Met.} \textbf{\bibinfo{volume}{103}},
  \bibinfo{pages}{2555} (\bibinfo{year}{1999}).

\bibitem[{\citenamefont{O'Connell et~al.}(2002)\citenamefont{O'Connell,
  Bachilo, Huffmann, Moore, Strano, Haroz, Rialon, Boul, Noon, Kittrell,
  Ma, Hauge, Weissman, and Smalley}}]{oconnell02}
\bibinfo{author}{\bibfnamefont{M.~J.} \bibnamefont{O'Connell}},
  \bibinfo{author}{\bibfnamefont{S.~M.} \bibnamefont{Bachilo}},
  \bibinfo{author}{\bibfnamefont{C.~B.} \bibnamefont{Huffmann}},
  \bibinfo{author}{\bibfnamefont{V.~C.} \bibnamefont{Moore}},
  \bibinfo{author}{\bibfnamefont{M.~S.} \bibnamefont{Strano}},
  \bibinfo{author}{\bibfnamefont{E.~H.} \bibnamefont{Haroz}},
  \bibinfo{author}{\bibfnamefont{K.~L.} \bibnamefont{Rialon}},
  \bibinfo{author}{\bibfnamefont{P.~J.} \bibnamefont{Boul}},
  \bibinfo{author}{\bibfnamefont{W.~H.} \bibnamefont{Noon}},
  \bibinfo{author}{\bibfnamefont{C.}~\bibnamefont{Kittrell}},
  \bibinfo{author}{\bibfnamefont{J.} \bibnamefont{Ma}},
  \bibinfo{author}{\bibfnamefont{R.~H.} \bibnamefont{Hauge}},
  \bibinfo{author}{\bibfnamefont{R.~B.}~\bibnamefont{Weissman}}, \bibnamefont{and}
  \bibinfo{author}{\bibfnamefont{R.~E.}~\bibnamefont{Smalley}},
  \bibinfo{journal}{Science}
  \textbf{\bibinfo{volume}{297}}, \bibinfo{pages}{593} (\bibinfo{year}{2002}).

\bibitem[{\citenamefont{Ugawa et~al.}(1999)\citenamefont{Ugawa, Rinzler, and
  Tanner}}]{ugawa99}
\bibinfo{author}{\bibfnamefont{A.}~\bibnamefont{Ugawa}},
  \bibinfo{author}{\bibfnamefont{A.~G.} \bibnamefont{Rinzler}},
  \bibnamefont{and} \bibinfo{author}{\bibfnamefont{D.~B.}
  \bibnamefont{Tanner}}, \bibinfo{journal}{Phys.\ Rev.\ B}
  \textbf{\bibinfo{volume}{60}}, \bibinfo{pages}{R11305}
  (\bibinfo{year}{1999}).

\bibitem[{\citenamefont{Ugawa et~al.}(2001{\natexlab{a}})\citenamefont{Ugawa,
  Hwang, Gommans, Tashiro, Rinzler, and Tanner}}]{ugawa01}
\bibinfo{author}{\bibfnamefont{A.}~\bibnamefont{Ugawa}},
  \bibinfo{author}{\bibfnamefont{J.}~\bibnamefont{Hwang}},
  \bibinfo{author}{\bibfnamefont{H.~H.} \bibnamefont{Gommans}},
  \bibinfo{author}{\bibfnamefont{H.}~\bibnamefont{Tashiro}},
  \bibinfo{author}{\bibfnamefont{A.~G.} \bibnamefont{Rinzler}},
  \bibnamefont{and} \bibinfo{author}{\bibfnamefont{D.~B.}
  \bibnamefont{Tanner}}, \bibinfo{journal}{Curr.\ Appl.\ Phys.}
  \textbf{\bibinfo{volume}{1}}, \bibinfo{pages}{45}
  (\bibinfo{year}{2001}{\natexlab{a}}).

\bibitem[{\citenamefont{Itkis et~al.}(2002)\citenamefont{Itkis, Niyogi, Meng,
  Hamon, Hu, and Haddon}}]{itkis02}
\bibinfo{author}{\bibfnamefont{M.~E.} \bibnamefont{Itkis}},
  \bibinfo{author}{\bibfnamefont{S.}~\bibnamefont{Niyogi}},
  \bibinfo{author}{\bibfnamefont{M.~E.} \bibnamefont{Meng}},
  \bibinfo{author}{\bibfnamefont{M.~A.} \bibnamefont{Hamon}},
  \bibinfo{author}{\bibfnamefont{H.}~\bibnamefont{Hu}}, \bibnamefont{and}
  \bibinfo{author}{\bibfnamefont{R.~C.} \bibnamefont{Haddon}},
  \bibinfo{journal}{Nano\ Lett.} \textbf{\bibinfo{volume}{2}},
  \bibinfo{pages}{155} (\bibinfo{year}{2002}).

\bibitem[{\citenamefont{Kane and Mele}(1997)}]{kane97}
\bibinfo{author}{\bibfnamefont{C.~L.} \bibnamefont{Kane}} \bibnamefont{and}
  \bibinfo{author}{\bibfnamefont{E.~J.} \bibnamefont{Mele}},
  \bibinfo{journal}{Phys.\ Rev.\ Lett.} \textbf{\bibinfo{volume}{78}},
  \bibinfo{pages}{1932} (\bibinfo{year}{1997}).

\bibitem[{\citenamefont{Delaney et~al.}(1998)\citenamefont{Delaney, Choi, Ihm,
  Louie, and Cohen}}]{delaney98}
\bibinfo{author}{\bibfnamefont{P.}~\bibnamefont{Delaney}},
  \bibinfo{author}{\bibfnamefont{H.~J.} \bibnamefont{Choi}},
  \bibinfo{author}{\bibfnamefont{J.}~\bibnamefont{Ihm}},
  \bibinfo{author}{\bibfnamefont{S.~G.} \bibnamefont{Louie}}, \bibnamefont{and}
  \bibinfo{author}{\bibfnamefont{M.~L.} \bibnamefont{Cohen}},
  \bibinfo{journal}{Nature} \textbf{\bibinfo{volume}{391}},
  \bibinfo{pages}{466} (\bibinfo{year}{1998}).

\bibitem[{\citenamefont{Maarouf et~al.}(2000)\citenamefont{Maarouf, Kane, and
  Mele}}]{maarouf00}
\bibinfo{author}{\bibfnamefont{A.~A.} \bibnamefont{Maarouf}},
  \bibinfo{author}{\bibfnamefont{C.~L.} \bibnamefont{Kane}}, \bibnamefont{and}
  \bibinfo{author}{\bibfnamefont{E.~J.} \bibnamefont{Mele}},
  \bibinfo{journal}{Phys.\ Rev.\ B} \textbf{\bibinfo{volume}{61}},
  \bibinfo{pages}{11156} (\bibinfo{year}{2000}).

\bibitem[{\citenamefont{Hwang et~al.}(2000)\citenamefont{Hwang, Gommans, Ugawa,
  Tashiro, Haggenmueller, Winey, Fischer, Tanner, and Rinzler}}]{hwang00}
\bibinfo{author}{\bibfnamefont{J.}~\bibnamefont{Hwang}},
  \bibinfo{author}{\bibfnamefont{H.~H.} \bibnamefont{Gommans}},
  \bibinfo{author}{\bibfnamefont{A.}~\bibnamefont{Ugawa}},
  \bibinfo{author}{\bibfnamefont{H.}~\bibnamefont{Tashiro}},
  \bibinfo{author}{\bibfnamefont{R.}~\bibnamefont{Haggenmueller}},
  \bibinfo{author}{\bibfnamefont{K.~I.} \bibnamefont{Winey}},
  \bibinfo{author}{\bibfnamefont{J.~E.} \bibnamefont{Fischer}},
  \bibinfo{author}{\bibfnamefont{D.~B.} \bibnamefont{Tanner}},
  \bibnamefont{and} \bibinfo{author}{\bibfnamefont{A.~G.}
  \bibnamefont{Rinzler}}, \bibinfo{journal}{Phys.\ Rev.\ B}
  \textbf{\bibinfo{volume}{62}}, \bibinfo{pages}{R13310}
  (\bibinfo{year}{2000}).

\bibitem[{\citenamefont{Hennrich et~al.}(2002)\citenamefont{Hennrich, Lebedkin,
  Malik, Tracy, Barczewski, {H. R\"osner}, and Kappes}}]{hennrich02}
\bibinfo{author}{\bibfnamefont{F.}~\bibnamefont{Hennrich}},
  \bibinfo{author}{\bibfnamefont{S.}~\bibnamefont{Lebedkin}},
  \bibinfo{author}{\bibfnamefont{S.}~\bibnamefont{Malik}},
  \bibinfo{author}{\bibfnamefont{J.}~\bibnamefont{Tracy}},
  \bibinfo{author}{\bibfnamefont{M.}~\bibnamefont{Barczewski}},
  \bibinfo{author}{\bibnamefont{{H. R\"osner}}}, \bibnamefont{and}
  \bibinfo{author}{\bibfnamefont{M.}~\bibnamefont{Kappes}},
  \bibinfo{journal}{Phys.\ Chem.\ Chem.\ Phys.} \textbf{\bibinfo{volume}{4}},
  \bibinfo{pages}{2273} (\bibinfo{year}{2002}).

\bibitem[{\citenamefont{Hennrich et~al.}(2003)\citenamefont{Hennrich, Wellmann,
  Malik, Lebedkin, and Kappes}}]{hennrich03}
\bibinfo{author}{\bibfnamefont{F.}~\bibnamefont{Hennrich}},
  \bibinfo{author}{\bibfnamefont{R.}~\bibnamefont{Wellmann}},
  \bibinfo{author}{\bibfnamefont{S.}~\bibnamefont{Malik}},
  \bibinfo{author}{\bibfnamefont{S.}~\bibnamefont{Lebedkin}}, \bibnamefont{and}
  \bibinfo{author}{\bibfnamefont{M.~M.} \bibnamefont{Kappes}},
  \bibinfo{journal}{Phys.\ Chem.\ Chem.\ Phys.} \textbf{\bibinfo{volume}{5}},
  \bibinfo{pages}{178} (\bibinfo{year}{2003}).

\bibitem[{\citenamefont{Rinzler et~al.}(1998)\citenamefont{Rinzler, Liu, Dai,
  Nikolaev, Huffmann, {F.~J. Rodriguez-Mac\'{\i}as}, Boul, Lu, Heymann, Colbert,
  Lee, Fischer, Rao, Eklund, and Smalley}}]{rinzler98}
\bibinfo{author}{\bibfnamefont{A.~G.} \bibnamefont{Rinzler}},
  \bibinfo{author}{\bibfnamefont{J.}~\bibnamefont{Liu}},
  \bibinfo{author}{\bibfnamefont{H.}~\bibnamefont{Dai}},
  \bibinfo{author}{\bibfnamefont{P.}~\bibnamefont{Nikolaev}},
  \bibinfo{author}{\bibfnamefont{C.~B.} \bibnamefont{Huffmann}},
  \bibinfo{author}{\bibnamefont{{F.~J. Rodriguez-Mac\'{\i}as}}},
  \bibinfo{author}{\bibfnamefont{P.~J.} \bibnamefont{Boul}},
  \bibinfo{author}{\bibfnamefont{A.~H.} \bibnamefont{Lu}},
  \bibinfo{author}{\bibfnamefont{D.}~\bibnamefont{Heymann}},
  \bibinfo{author}{\bibfnamefont{D.~T.} \bibnamefont{Colbert}},
  \bibinfo{author}{\bibnamefont{{R.~S. Lee}}},
  \bibinfo{author}{\bibfnamefont{J.~E.} \bibnamefont{Fischer}},
  \bibinfo{author}{\bibfnamefont{A.~M.} \bibnamefont{Rao}},
  \bibinfo{author}{\bibfnamefont{P.~C.}~\bibnamefont{Eklund}}, \bibnamefont{and}
  \bibinfo{author}{\bibfnamefont{R.~E.} \bibnamefont{Smalley}},
  \bibinfo{journal}{Appl.\ Phys.\ A}
  \textbf{\bibinfo{volume}{67}}, \bibinfo{pages}{29} (\bibinfo{year}{1998}).

\bibitem[{\citenamefont{Dressel and Gr\"uner}(2002)}]{dresselgruner02}
\bibinfo{author}{\bibfnamefont{M.}~\bibnamefont{Dressel}} \bibnamefont{and}
  \bibinfo{author}{\bibfnamefont{G.}~\bibnamefont{Gr\"uner}},
  \emph{\bibinfo{title}{Electrodynamics of solids}}
  (\bibinfo{publisher}{Cambridge University Press}, \bibinfo{year}{2002}).

\bibitem[{\citenamefont{Zhao et~al.}(2004)\citenamefont{Zhao, Itkis, Niyogi,
  Hu, Zhang, and Haddon}}]{zhao04}
\bibinfo{author}{\bibfnamefont{B.}~\bibnamefont{Zhao}},
  \bibinfo{author}{\bibfnamefont{M.~E.} \bibnamefont{Itkis}},
  \bibinfo{author}{\bibfnamefont{S.}~\bibnamefont{Niyogi}},
  \bibinfo{author}{\bibfnamefont{H.}~\bibnamefont{Hu}},
  \bibinfo{author}{\bibfnamefont{J.}~\bibnamefont{Zhang}}, \bibnamefont{and}
  \bibinfo{author}{\bibfnamefont{R.~C.} \bibnamefont{Haddon}},
  \bibinfo{journal}{J.\ Phys.\ Chem.\ B} \textbf{\bibinfo{volume}{108}},
  \bibinfo{pages}{8136} (\bibinfo{year}{2004}).

\bibitem[{\citenamefont{Kim et~al.}(2005)\citenamefont{Kim, Liu, Furtado, Chen,
  Saito, Jiang, Dresselhaus, and Eklund}}]{Kim05}
\bibinfo{author}{\bibfnamefont{U.~J.} \bibnamefont{Kim}},
  \bibinfo{author}{\bibfnamefont{X.~M.} \bibnamefont{Liu}},
  \bibinfo{author}{\bibfnamefont{C.}~\bibnamefont{Furtado}},
  \bibinfo{author}{\bibfnamefont{G.}~\bibnamefont{Chen}},
  \bibinfo{author}{\bibfnamefont{R.}~\bibnamefont{Saito}},
  \bibinfo{author}{\bibfnamefont{J.}~\bibnamefont{Jiang}},
  \bibinfo{author}{\bibfnamefont{M.~S.} \bibnamefont{Dresselhaus}},
  \bibnamefont{and} \bibinfo{author}{\bibfnamefont{P.~C.}
  \bibnamefont{Eklund}}, \bibinfo{journal}{Phys.\ Rev.\ Lett.}
  \textbf{\bibinfo{volume}{95}}, \bibinfo{pages}{157402}
  (\bibinfo{year}{2005}).

\bibitem[{\citenamefont{Murakami et~al.}(2005)\citenamefont{Murakami,
  Einarsson, Edamura, and Maruyama}}]{murakami05prl}
\bibinfo{author}{\bibfnamefont{Y.}~\bibnamefont{Murakami}},
  \bibinfo{author}{\bibfnamefont{E.}~\bibnamefont{Einarsson}},
  \bibinfo{author}{\bibfnamefont{T.}~\bibnamefont{Edamura}}, \bibnamefont{and}
  \bibinfo{author}{\bibfnamefont{S.}~\bibnamefont{Maruyama}},
  \bibinfo{journal}{Phys.\ Rev.\ Lett.} \textbf{\bibinfo{volume}{94}},
  \bibinfo{pages}{087402} (\bibinfo{year}{2005}).

\bibitem[{\citenamefont{Itkis et~al.}(2003)\citenamefont{Itkis, Perea, Niyogi,
  Rickard, Hamon, Hu, Zhao, and Haddon}}]{itkis03}
\bibinfo{author}{\bibfnamefont{M.~E.} \bibnamefont{Itkis}},
  \bibinfo{author}{\bibfnamefont{D.~E.} \bibnamefont{Perea}},
  \bibinfo{author}{\bibfnamefont{S.}~\bibnamefont{Niyogi}},
  \bibinfo{author}{\bibfnamefont{S.~M.} \bibnamefont{Rickard}},
  \bibinfo{author}{\bibfnamefont{M.~A.} \bibnamefont{Hamon}},
  \bibinfo{author}{\bibfnamefont{H.}~\bibnamefont{Hu}},
  \bibinfo{author}{\bibfnamefont{B.}~\bibnamefont{Zhao}}, \bibnamefont{and}
  \bibinfo{author}{\bibfnamefont{R.~C.} \bibnamefont{Haddon}},
  \bibinfo{journal}{Nano\ Lett.} \textbf{\bibinfo{volume}{3}},
  \bibinfo{pages}{309} (\bibinfo{year}{2003}).

\bibitem[{\citenamefont{Jeon et~al.}(2004)\citenamefont{Jeon, Kim, Kang, Maeng,
  Son, An, Lee, and Lee}}]{jeon04}
\bibinfo{author}{\bibfnamefont{T.-I.} \bibnamefont{Jeon}},
  \bibinfo{author}{\bibfnamefont{K.-J.} \bibnamefont{Kim}},
  \bibinfo{author}{\bibfnamefont{C.}~\bibnamefont{Kang}},
  \bibinfo{author}{\bibfnamefont{I.~H.} \bibnamefont{Maeng}},
  \bibinfo{author}{\bibfnamefont{J.-H.} \bibnamefont{Son}},
  \bibinfo{author}{\bibfnamefont{K.~H.} \bibnamefont{An}},
  \bibinfo{author}{\bibfnamefont{J.~Y.} \bibnamefont{Lee}}, \bibnamefont{and}
  \bibinfo{author}{\bibfnamefont{Y.~H.} \bibnamefont{Lee}},
  \bibinfo{journal}{J.\ Appl.\ Phys.} \textbf{\bibinfo{volume}{95}},
  \bibinfo{pages}{5736} (\bibinfo{year}{2004}).

\bibitem[{\citenamefont{Hilt et~al.}(2000)\citenamefont{Hilt, Brom, Ugawa, and
  Ahlskog}}]{hilt00}
\bibinfo{author}{\bibfnamefont{O.}~\bibnamefont{Hilt}},
  \bibinfo{author}{\bibfnamefont{H.~B.} \bibnamefont{Brom}},
  \bibinfo{author}{\bibfnamefont{A.}~\bibnamefont{Ugawa}}, \bibnamefont{and}
  \bibinfo{author}{\bibfnamefont{M.}~\bibnamefont{Ahlskog}},
  \bibinfo{journal}{Phys.\ Rev.\ B} \textbf{\bibinfo{volume}{61}},
  \bibinfo{pages}{R5130} (\bibinfo{year}{2000}).

\bibitem[{\citenamefont{Ouyang et~al.}(2001)\citenamefont{Ouyang, Huang,
  Cheung, and Lieber}}]{ouyang01}
\bibinfo{author}{\bibfnamefont{M.}~\bibnamefont{Ouyang}},
  \bibinfo{author}{\bibfnamefont{J.-L.} \bibnamefont{Huang}},
  \bibinfo{author}{\bibfnamefont{C.~L.} \bibnamefont{Cheung}},
  \bibnamefont{and} \bibinfo{author}{\bibfnamefont{C.~M.}
  \bibnamefont{Lieber}}, \bibinfo{journal}{Science}
  \textbf{\bibinfo{volume}{292}}, \bibinfo{pages}{702} (\bibinfo{year}{2001}).

\bibitem[{\citenamefont{Ugawa et~al.}(2001{\natexlab{b}})\citenamefont{Ugawa,
  Rinzler, and Tanner}}]{ugawafer01}
\bibinfo{author}{\bibfnamefont{A.}~\bibnamefont{Ugawa}},
  \bibinfo{author}{\bibfnamefont{A.~G.} \bibnamefont{Rinzler}},
  \bibnamefont{and} \bibinfo{author}{\bibfnamefont{D.~B.}
  \bibnamefont{Tanner}}, \bibinfo{journal}{Ferroelectrics}
  \textbf{\bibinfo{volume}{249}}, \bibinfo{pages}{145}
  (\bibinfo{year}{2001}{\natexlab{b}}).

\bibitem[{\citenamefont{Sumanasekera et~al.}(2000)\citenamefont{Sumanasekera,
  Adu, Fang, and Eklund}}]{sumana00}
\bibinfo{author}{\bibfnamefont{G.~U.} \bibnamefont{Sumanasekera}},
  \bibinfo{author}{\bibfnamefont{C.~K.~W.} \bibnamefont{Adu}},
  \bibinfo{author}{\bibfnamefont{S.}~\bibnamefont{Fang}}, \bibnamefont{and}
  \bibinfo{author}{\bibfnamefont{P.~C.} \bibnamefont{Eklund}},
  \bibinfo{journal}{Phys.\ Rev.\ Lett.} \textbf{\bibinfo{volume}{85}},
  \bibinfo{pages}{1096} (\bibinfo{year}{2000}).

\bibitem[{\citenamefont{Collins et~al.}(2000)\citenamefont{Collins, Bradley,
  Ishigami, and Zettl}}]{collins00}
\bibinfo{author}{\bibfnamefont{P.~G.} \bibnamefont{Collins}},
  \bibinfo{author}{\bibfnamefont{K.}~\bibnamefont{Bradley}},
  \bibinfo{author}{\bibfnamefont{M.}~\bibnamefont{Ishigami}}, \bibnamefont{and}
  \bibinfo{author}{\bibfnamefont{A.}~\bibnamefont{Zettl}},
  \bibinfo{journal}{Science} \textbf{\bibinfo{volume}{287}},
  \bibinfo{pages}{1801} (\bibinfo{year}{2000}).

\bibitem[{\citenamefont{Ishii et~al.}(2003)\citenamefont{Ishii, Kataura,
  Shiozawa, Yoshioka, Otsubo, Takayama, Miyahara, Suzuki, Achiba, Nakataka,
  Narimura, Higashiguchi, Shimada, Namatame, and Taniguchi}}]{ishii03}
\bibinfo{author}{\bibfnamefont{H.}~\bibnamefont{Ishii}},
  \bibinfo{author}{\bibfnamefont{H.}~\bibnamefont{Kataura}},
  \bibinfo{author}{\bibfnamefont{H.}~\bibnamefont{Shiozawa}},
  \bibinfo{author}{\bibfnamefont{H.}~\bibnamefont{Yoshioka}},
  \bibinfo{author}{\bibfnamefont{H.}~\bibnamefont{Otsubo}},
  \bibinfo{author}{\bibfnamefont{Y.}~\bibnamefont{Takayama}},
  \bibinfo{author}{\bibfnamefont{T.}~\bibnamefont{Miyahara}},
  \bibinfo{author}{\bibfnamefont{S.}~\bibnamefont{Suzuki}},
  \bibinfo{author}{\bibfnamefont{Y.}~\bibnamefont{Achiba}},
  \bibinfo{author}{\bibfnamefont{M.}~\bibnamefont{Nakataka}},
  \bibinfo{author}{\bibfnamefont{T.}~\bibnamefont{Narimura}},
  \bibinfo{author}{\bibfnamefont{M.}~\bibnamefont{Higashiguchi}},
  \bibinfo{author}{\bibfnamefont{K.}~\bibnamefont{Shimada}},
  \bibinfo{author}{\bibfnamefont{H.}~\bibnamefont{Namatame}}, \bibnamefont{and}
  \bibinfo{author}{\bibfnamefont{M.}~\bibnamefont{Taniguchi}},
  \bibinfo{journal}{Nature}
  \textbf{\bibinfo{volume}{426}}, \bibinfo{pages}{540} (\bibinfo{year}{2003}).

\bibitem[{\citenamefont{Liu et~al.}(2004)\citenamefont{Liu, Pichler, Knupfer,
  Fink, and Kataura}}]{liu04}
\bibinfo{author}{\bibfnamefont{X.}~\bibnamefont{Liu}},
  \bibinfo{author}{\bibfnamefont{T.}~\bibnamefont{Pichler}},
  \bibinfo{author}{\bibfnamefont{M.}~\bibnamefont{Knupfer}},
  \bibinfo{author}{\bibfnamefont{J.}~\bibnamefont{Fink}}, \bibnamefont{and}
  \bibinfo{author}{\bibfnamefont{H.}~\bibnamefont{Kataura}},
  \bibinfo{journal}{Phys.\ Rev.\ B} \textbf{\bibinfo{volume}{70}},
  \bibinfo{pages}{205405} (\bibinfo{year}{2004}).

\bibitem[{\citenamefont{Kazaoui et~al.}(1999)\citenamefont{Kazaoui, Minami,
  Jacquemin, Kataura, and Achiba}}]{kaza99}
\bibinfo{author}{\bibfnamefont{S.}~\bibnamefont{Kazaoui}},
  \bibinfo{author}{\bibfnamefont{N.}~\bibnamefont{Minami}},
  \bibinfo{author}{\bibfnamefont{R.}~\bibnamefont{Jacquemin}},
  \bibinfo{author}{\bibfnamefont{H.}~\bibnamefont{Kataura}}, \bibnamefont{and}
  \bibinfo{author}{\bibfnamefont{Y.}~\bibnamefont{Achiba}},
  \bibinfo{journal}{Phys.\ Rev.\ B} \textbf{\bibinfo{volume}{60}},
  \bibinfo{pages}{13339} (\bibinfo{year}{1999}).

\bibitem[{\citenamefont{Petit et~al.}(1999)\citenamefont{Petit, Mathis,
  Journet, and Bernier}}]{petit99}
\bibinfo{author}{\bibfnamefont{P.}~\bibnamefont{Petit}},
  \bibinfo{author}{\bibfnamefont{C.}~\bibnamefont{Mathis}},
  \bibinfo{author}{\bibfnamefont{C.}~\bibnamefont{Journet}}, \bibnamefont{and}
  \bibinfo{author}{\bibfnamefont{P.}~\bibnamefont{Bernier}},
  \bibinfo{journal}{Chem. Phys. Lett.} \textbf{\bibinfo{volume}{305}},
  \bibinfo{pages}{370} (\bibinfo{year}{1999}).

\bibitem[{\citenamefont{Pichler et~al.}(1999)\citenamefont{Pichler, Sing,
  Knupfer, Golden, and Fink}}]{pichler99}
\bibinfo{author}{\bibfnamefont{T.}~\bibnamefont{Pichler}},
  \bibinfo{author}{\bibfnamefont{M.}~\bibnamefont{Sing}},
  \bibinfo{author}{\bibfnamefont{M.}~\bibnamefont{Knupfer}},
  \bibinfo{author}{\bibfnamefont{M.~S.} \bibnamefont{Golden}},
  \bibnamefont{and} \bibinfo{author}{\bibfnamefont{J.}~\bibnamefont{Fink}},
  \bibinfo{journal}{Solid State Commun.} \textbf{\bibinfo{volume}{109}},
  \bibinfo{pages}{721} (\bibinfo{year}{1999}).

\end{thebibliography}

\end{document}